\documentstyle[aps,prc,twocolumn,epsfig]{revtex}
\begin{document}
%\draft
 
\title{Collectivity Embedded in Complex Spectra:\\
Example of Nuclear Double-Charge Exchange Modes}
 
\author{S. Dro\.zd\.z$^{1,2}$, S. Nishizaki$^{3}$, J. Speth$^{1}$
and M. W\'ojcik$^{1,2}$}
\address{
$^{1}$Institut f\"ur Kernphysik, Forschungszentrum J\"ulich,\\
D-52425 J\"ulich, Germany \\
$^{2}$ Institute of Nuclear Physics, PL-31-342 Krak\'ow, Poland\\
$^{3}$ Faculty of Humanities and Social Sciences, Iwate University, \\
Ueda 3-18-34, Morioka 020, Japan }
\date{\today}
\maketitle

\begin{abstract}
The mechanism of collectivity coexisting with chaos is investigated 
on the quantum level. The complex spectra are represented in the basis
of two-particle two-hole states describing the nuclear  
double-charge exchange modes in $^{48}$Ca.
An example of $J^{\pi}=0^-$ excitations shows that the residual interaction, 
which generically implies chaotic behavior, 
under certain specific and well identified conditions may create transitions
stronger than those corresponding to the pure mean-field picture.
Therefore, for this type of excitations such an effect is not generic 
and in most cases the strength of transitions is likely to take much lower 
values, even close to the Porter-Thomas distributed.    
\end{abstract}
 
 
\bigskip
 
%\newpage

Chaos is essentially a generic property of complex systems such as atomic
nuclei and this finds evidence in a broad applicability of the 
random matrix theory (RMT)~\cite{Brod} to discribe level fluctuations. 
Another characteristics connected with complexity,
even more interesting and important from the practical point of view,
is collectivity. It means a cooperation, and thus the coupling, between
the different degrees of freedom in order to generate a coherent signal
in response to an external perturbation. Consequently, even though the
real collectivity implies a highly ordered behavior it involves effects
beyond the mean field -- the most regular part of the
nuclear Hamiltonian~\cite{Zele}. At the same time the effects beyond the
mean field are responsible  for the fluctuation properties characteristic of
the Gaussian orthogonal ensemble (GOE) of random matrices~\cite{ZBFH}.
 
Local level fluctuations characteristic of GOE appear~\cite{Dro1}
to take place for the nuclear Hamiltonian acting already in the space
of two-particle -- two-hole (2p2h) states and this is a crucial
element for an appropriate description of the giant resonance decay
properties~\cite{Dro2}. The ordinary giant resonances are, however, excited by
one-body operators which directly probe
the 1p1h components of the nuclear wave function.
The 2p2h states only form the background which determines a decay-law.
There exist, however, very interesting physical processes,
represented by two-body external operators, which directly couple
the ground state to the space of 2p2h states.
In view of the above mentioned local GOE fluctuations giving evidence
for a significant amount of
chaotic dynamics already in the 2p2h space, the question of a possible
coherent response under such conditions
is a very intriguing one and of interest not only for many branches of physics 
but also for biological systems~\cite{Kauf}.
 
Among various nuclear excitation modes which can be considered in this
context the double charge exchange (DCX) processes are of special
interest. These modes, excited in $({\pi}^+,{\pi}^-)$
reactions~\cite{Leit},
involve at least two nucleons within the nucleus and give rise to
a sharp peak at around 50 MeV in the forward cross section.
They are thus located in the energy region of the high density of 2p2h
states which points to the importance of coherence effects among those
states. Consequently, the present investigation may also appear helpful
in future studies of the mechanism of DCX reactions and in separating
the suggested~\cite{Bilg} dibaryon contribution from the conventional
effects~\cite{Kaga}.
 
Diagonalizing the nuclear Hamiltonian in the subspace of 2p2h states
$\vert 2 \rangle \equiv
a_{p_1}^{\dag} a_{p_2}^{\dag}  a_{h_2}a_{h_1}\vert 0 \rangle$
yields the eigenenergies $E_n$ and the corresponding eigenvectors
$\vert n \rangle = \Sigma_2 c^n_2 \vert 2 \rangle$.
In response to an external field ${\hat F}_{\alpha}$ a state
%\begin{equation}
$\vert F_{\alpha} \rangle \equiv  {\hat F}_{\alpha} \vert 0 \rangle =
\sum_n \langle n \vert {\hat F}_{\alpha} \vert 0 \rangle \vert n \rangle$
%\label{eq:F}
%\end{equation}
is excited. The two-phonon operator ${\hat F}_{\alpha}$ can be
represented as ${\hat F}_{\alpha}=
\{{\hat f}_{\beta}\otimes{\hat f}_{\gamma}\}_{\alpha}$,
where ${\hat f}_{\beta}$ and ${\hat f}_{\gamma}$ denote the single-phonon
operators whose quantum numbers $\beta$ and $\gamma$ are coupled
to form $\alpha$.
The state $\vert F_{\alpha} \rangle$ determines the strength function
\begin{equation}
S_{F_{\alpha}}(E) = \sum_n S_{F_{\alpha}}(n) \delta (E-E_n),
\label{eq:S}
\end{equation}
where $S_{F_{\alpha}}(n) =
\vert\langle n \vert{\hat F}_{\alpha}\vert 0 \rangle\vert^2$.
In the basis of states $\vert 2 \rangle$
\begin{equation}
S_{F_{\alpha}}(n)
= \sum_2 \vert c^n_2 \vert^2
\vert \langle 2 \vert {\hat F}_{\alpha} \vert 0 \rangle \vert^2
+ \sum_{2\neq2'} c^{n*}_2 c^n_{2'}
\langle 0 \vert {\hat F}_{\alpha}^{\dag} \vert 2'\rangle
\langle 2 \vert {\hat F}_{\alpha} \vert 0 \rangle.
\label{eq:Sn}
\end{equation}
This equality defines its diagonal $(S^d_{F_{\alpha}}(n))$ and off-diagonal
$(S^{od}_{F_{\alpha}}(n))$ contributions.
The second component includes many more terms
and it is this component which potentially is able to induce
collectivity, i.e. a strong transition to energy $E_n$.
Two elements are however required: (i) a state
$\vert n \rangle$ must involve sufficiently many
expansion coefficients $c^n_2$ over the unperturbed states
$\vert 2 \rangle$ which carry the strength
($\langle 2 \vert {\hat F}_{\alpha} \vert 0 \rangle \neq 0$) and
this is equivalent to at least local mixing,
but at the same time (ii) sign correlations among these expansion
coefficients should take place so that the different terms do not cancel
out.
 
Optimal circumstances for the second condition to be fulfilled read:
$c^n_2 \sim \langle 0 \vert {\hat F}_{\alpha} \vert 2 \rangle$.
This may occur if the interaction matrix elements can be represented by a sum
of separable terms ${\bf Q}^{\nu}$ of the multipole-multipole type 
(${\cal V}_{ij,kl}=\sum_{\nu=1}^M Q^{\nu}_{ij} Q^{\nu}_{kl}$
with $Q^{\nu}_{ij} \sim \langle i \vert {\hat f}_{\nu} \vert j \rangle$). 
The success of the Brown-Bolsterli schematic model~\cite{BB} 
in indicating the mechanism of collectivity on the 1p1h level
points to an approximate validity of such a representation.
Collectivity can then be viewed as an edge effect connected with 
appearance of a dominating component in the Hamiltonian matrix and the rank 
$M$ of this component is significantly lower (unity in case of the 
Brown-Bolsterli model) than the size of the matrix. This rank specifies 
a number of the prevailing states whose expansion coefficients predominantly 
are functions of ${\bf Q}^{\nu}$.  
Due to two-body nature of the nuclear interaction which reduces its 2p2h
matrix elements to combinations of the ones representing the 
particle-particle, hole-hole and particle-hole interactions~\cite{Dro1}
the separability may become effective also on the 2p2h level. 
 
For quantitative discussion presented below we choose the $^{48}$Ca
nucleus, specify the mean field part of the Hamiltonian
in terms of a local Woods-Saxon potential including the Coulomb
interaction and adopt the density-dependent zero-range interaction
of ref.~\cite{SW} as a residual interaction.
Since we want to inspect the higher energy region
at least three mean field shells on both sides
of the Fermi surface have to be used to generate the unperturbed 2p2h
states as a basis for diagonalization of the full Hamiltonian.
Typically, the number of such states is very large and this kind of
calculation can be kept under full numerical control only for excitations
of the lowest multipolarity. For this reason
we perform a systematic study of the DCX $J^{\pi}=0^-$ states.
Our model space then develops N=2286 2p2h states.
There are still several possibilities of exciting such a double-phonon
mode represented by the operator ${\hat F}_{\alpha}$
out of the two single-phonons ${\hat f}_{\beta}$ and ${\hat f}_{\gamma}$
of opposite parity. For definitness we choose
${\hat f}_{\beta}= r Y_1 {\tau}_-$ and
${\hat f}_{\gamma}= r^2 [Y_2 \otimes \sigma]^{1^+} {\tau}_-$.
The first of these operators corresponds to the $1\hbar\omega$ dipole
and the second to $2\hbar\omega$ spin-quadrupole excitation.
The resulting two-phonon mode thus operates on a level of
$3\hbar\omega$ excitations.

The results of calculations are presented in Fig.~1. As one can see,
including the residual interaction (part (b)) induces a spectacularly
strong transition at 49.1 MeV.
This transition
is stronger by more than a factor of 2 than any of the unperturbed
(part (a)) transitions even though it is shifted to significantly
higher ($\sim$ 10 MeV) energy.
This is also a very collective transition. About $96\%$ of the
corresponding strength originates from $S^{od}_{F_{\alpha}}(n)$,
as comparison between parts (b) and (c) of Fig.~1 documents.
The degree of mixing can be quantified, for instance,
in terms of the information entropy~\cite{Izra}
$I(n)=-\Sigma_i p_i \ln p_i$ $(p_i=\vert c^n_i \vert^2)$ of an eigenvector
$\vert n \rangle$ in the basis (part (d)).
Interestingly, the system finds preferential conditions
for creating the most collective state in the energy region
of local minimum in $I(n)$.
Our following discussion is supposed to shead more light on this issue. 
%--------------------------------------------------------------------------
\begin{figure}[t]
\hspace{\fill}
%\begin{minipage}[t]{75mm}
\begin{center}
\epsfig{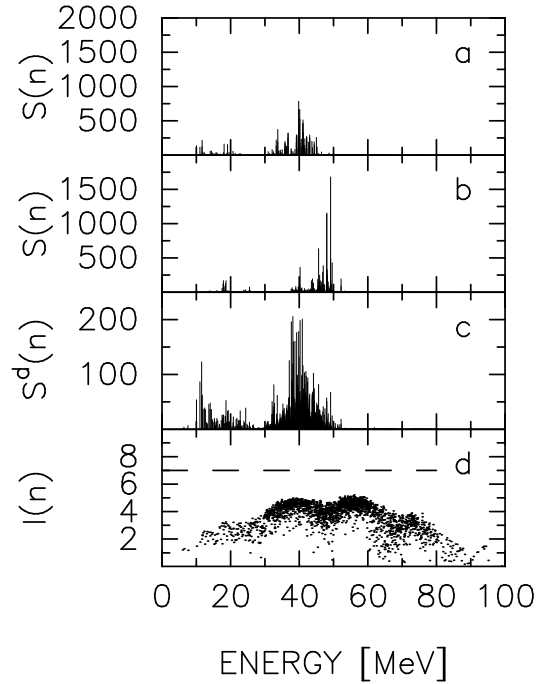}
\end{center}
%\end{minipage}
\caption{
(a) The unperturbed transition-strength distribution
in $^{48}Ca$ for the $J^{\pi} = 0^-$ DCX excitation involving
the single-phonon dipole and $2\hbar\omega$ spin-quadrupole modes.
(b) The same as (a) but after including the residual interaction.
(c) $S^d_{F_{\alpha}}$ components of the transition-strength as defined
by eq.~(\ref{eq:Sn}) (notice a different scale).
(d) The information entropy of the states 
$\vert n \rangle$ in the unperturbed basis. The dashed line indicates 
the GOE limit $(\ln(0.48N))$.
}
\end{figure}
%-------------------------------------------------------------------------- 
As shown in Fig.~2(a) our Hamiltonian matrix displays a band-like structure 
with spots of the significant matrix elements inside. This together with
the nonuniform energy distribution $\rho_u(E)$ of the unperturbed 2p2h states,
(Fig.~2(b)) which is a trace of the shell structure 
of the single particle states,
sizably suppresses the range of mixing and locally supports conditions for the
edge effect to occur in the energy region of the minimum in $\rho_u(E)$.
A comparison with Fig.~1(b) shows that the collective state is located 
at about this region. Moreover, the minimum survives diagonalization
($\rho_p(E)$ in Fig.~2(c)) and all the above features are consistent with
the effective band range~\cite{FLW} 
%\begin{equation}
($(\Delta E_i)^2 = \sum_j(H_{ii}-H_{jj})^2 H_{ij}^2 / \sum_{i \ne j} H_{ij}^2$)
%\label{eq:band}
%\end{equation}
shown in Fig.~2(d).

Further quantification of the character of mixing between the unperturbed 
states is documented in Fig.~3. The distribution of off-diagonal matrix
elements (a) is of the Porter-Thomas (P-T) -- type which indicates 
the presence of the dominating multipole-multipole components 
in the interaction~\cite{ZBFH,FGGK}. 
%----------------------------------------------------------------------------
\begin{figure}[t]
\hspace{\fill}
%\begin{minipage}[t]{75mm}
\begin{center}
\epsfig{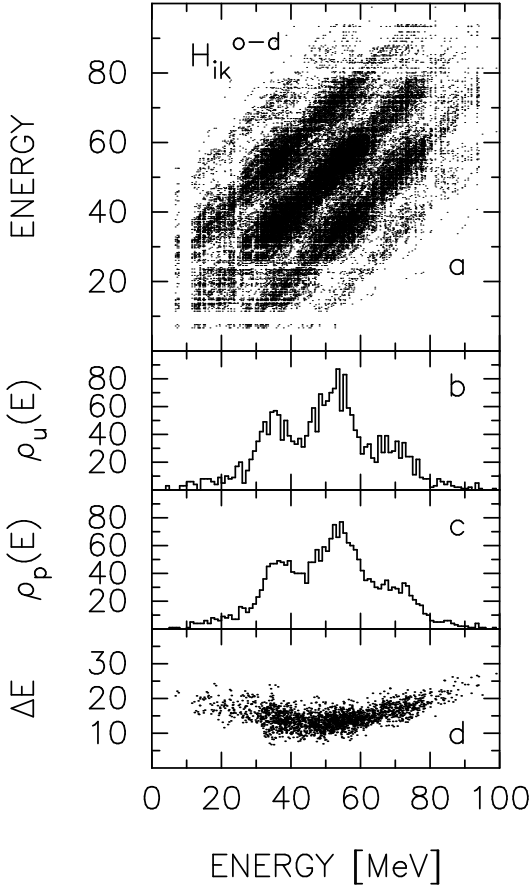}
\end{center}
%\end{minipage}
\caption{
(a) The structure of the Hamiltonian matrix for the 
$J^{\pi}=0^-$ DCX states. The states are here labeled by energies, 
ordered in ascending order and the matrix elements $H_{ik} \ge 0.1$
are indicated by the dots.
(b) The density of the unperturbed 2p2h-states.
(c) The density of states after the diagonalization.
(d) The energy range of interaction between the unperturbed states.\\
}
\end{figure}
%----------------------------------------------------------------------------
An interesting novel feature is the asymmetry between the positive and negative
valued matrix elements (see parameters in the caption to Fig.~3).
The positive matrix elements are more abundant 
which expresses further correlations among them
and the fact that the interaction is predominantly repulsive for the mode
considered.
As a chaos related characteristics we take the spectral rigidity measured
in terms of the $\Delta_3$ statistics~\cite{Brod}. We find this measure more
appropriate for studying various local subtleties of mixing than the 
nearest neighbor spacing (NNS) distribution because for a smaller  
number of states the latter 
sooner becomes contaminated by strong fluctuations.
Indeed, the spectral rigidity (Fig.~3(b)) detects differences in the level
repulsion inside the string of eigenvalues (i1) covering the first maximum in
$\rho_{p(u)}(E)$ (35.0-41.75 MeV) and the one (i2) covering the minimum 
and thus including the collective state (41.75-50.1 MeV). 
Again, the deviation from GOE is more significant in i2 which,
similarly as the behavior of $I(n)$, signals 
a more regular dynamics in the vicinity of the collective state $(n=996)$.     
%--------------------------------------------------------------------------------
\begin{figure}[t]
\hspace{\fill}
%\begin{minipage}[t]{75mm}
\begin{center}
\epsfig{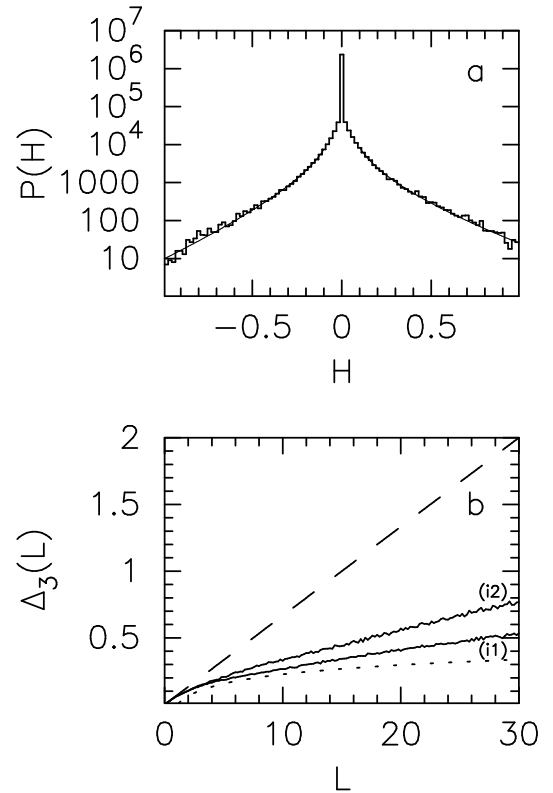}
\end{center}
%\end{minipage}
\caption{
(a) The distribution of off-diagonal matrix elements between
the $J^{\pi}=0^-$ DCX states (histogram). The solid lines indicate fit in terms
of $P(H)=a \vert H\vert^b \exp(-\vert H\vert/c)$ with the resulting parameters:
$a=683$, $b=-1.22$, $c=0.23$ (left) and $a=538$, $b=-1.32$, $c=0.32$ (right). 
(b) The spectral rigidity $\Delta_3(L)$ for eigenvalues from the 
two intervals: $n=351-700$ (i1) and $n=701-1050$ (i2).
The long-dashed line corresponds to Poisson level distribution and the
short-dashed line to GOE.\\
}
\end{figure}
%--------------------------------------------------------------------------------
The conditions corresponding to the actual nuclear Hamiltonian are not the
most optimal ones from the point of view of the collectivity of our 
$J^{\pi}=0^-$ DCX excitation. By multiplying the off-diagonal matrix elements
by a factor of 0.7 we obtain a picture as shown in Fig.~4(a and b). Now the
transition located at 46 MeV is another factor of 2 stronger than before.
However, the range of values of a
multiplication factor which produces this kind of picture is rather narrow
and this feature of collectivity resembles a classical
phenomenon of the stochastic resonance~\cite{Wies}. 
It is relatively easy to completely destroy such a strong transition. 
By multiplying 
the off-diagonal matrix elements by a factor of 3 
(which is equivalent to increasing the density of states) 
the strength distribution displays a form as shown in Fig.~4(c). 
%-----------------------------------------------------------------------------
\begin{figure}[t]
\hspace{\fill}
%\begin{minipage}[t]{75mm}
\begin{center}
\epsfig{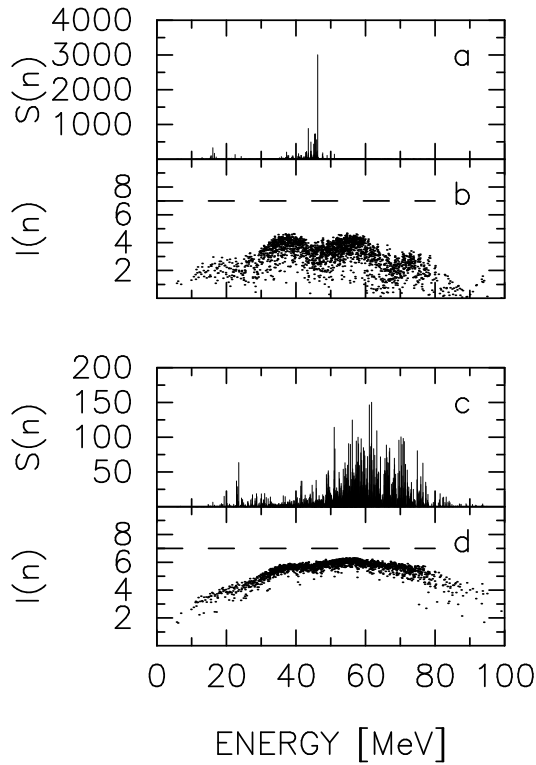}
\end{center}
%\end{minipage}
\caption{
The transition-strength distribution and the information entropy
for the same excitation as in Fig.~1 but off-diagonal matrix elements are now
multiplied by a factor of 0.7 ((a) and (b)) and by a factor of 3.0 
((c) and (d)).
}
\end{figure}
%------------------------------------------------------------------------------
Even though this strength remains largely 
localized in energy (standard way of analysing experimental data 
may even classify it as collective) the corresponding $P(S(n))$ 
is essentially P-T distributed (Gaussian distributed amplitudes)
which becomes evident from the entropy estimate. 
Normalizing the total strength to unity, identifying the
so normalized $S(n)$ with the probability $\rho_n$ to populate a state 
at energy $E_n$, and defining the corresponding total entropy  
${\cal I} = -\sum_n \rho_n \ln \rho_n$ one obtains ${\cal I}=6.82$ while
the P-T distribution for the same number of states $(N=2286)$
gives ${\cal I}=7.0$. Further increase of the multiplication factor
may again produce some transitions which are more collective than those 
allowed by P-T. 
In particular, starting from values $\sim$5 some new strong collective 
transitions appear at the upper edge of the whole spectrum but this is the
effect of basis truncation.
     
In conclusion, a real collectivity, by which we mean a transition stronger
than those generated by the mean field, is a very subtle effect and is not 
a generic property of the complex spectra. Its appearance, as it happens 
for one of the components of the $J^{\pi}=0^-$ DCX excitations considered 
here, involves several
elements like correlations among the matrix elements, nonuniformities in the
distribution of states and a proper matching of the interaction strength
to an initial (unperturbed) location of the transition strength relative 
to the scale of nonuniformities in the distribution of states.
If present, a collective state is then located in the region of more 
regular dynamics.  
This aspect of collectivity parallels an analogous
property hypothesised for living organisms~\cite{Kauf} and stating that
collectivity is a phenomenon occuring at the border 
between chaos and regularity.  
Based on the present study it is expected that majority of the two-phonon 
nuclear giant resonances is characterised by a spectrum of transitions
which are significantly smaller than those corresponding to the mean field 
picture and may even be compatible with the P-T distribution with the largest
transitions concentrated at similar energies. This latter effect may then 
lead to an illusion of collectivity.
 
We thank Franz Osterfeld for helpful discussions.
This work was supported in part by Polish KBN Grant No. 2 P03B 140 10.


\begin{references}

\bibitem{Brod} T.A.~Brody {\it et al.,}, Rev. Mod. Phys. {\bf 53}, 385(1981)
\bibitem{Zele} V.G.~Zelevinsky, Nucl. Phys. {\bf A555}, 109(1993)
\bibitem{ZBFH} V.G.~Zelevinsky, B.A.~Brown, N.~Frazier and M.~Horoi,
Phys. Rep. {\bf 276}, 85(1996)
\bibitem{Dro1} S.~Dro\.zd\.z, S.~Nishizaki, J.~Speth and J.~Wambach,
Phys. Rev. {\bf C49}, 867(1994)
\bibitem{Dro2} S.~Dro\.zd\.z, S.~Nishizaki and J.~Wambach,
Phys. Rev. Lett. {\bf 72}, 2839(1994)
\bibitem{Kauf} S.A.~Kauffman, in {\it The Principles of Organisation in
Organisms}, edited by J.E.~Mittenthal and A.B.~Baskin (Addion-Wesley, New York,
1992), p. 303 
\bibitem{Leit} M.J.~Leitch {\it et al.}, Phys. Rev. {\bf C39},
2356(1989)
\bibitem{Bilg} R.~Bilger, H.A.~Clement and M.G.~Schepkin,
Phys. Rev. Lett. {\bf 71}, 42(1993)
\bibitem{Kaga} M.A.~Kagarlis and M.B.~Johnson,
Phys. Rev. Lett. {\bf 73}, 38(1994)
\bibitem{BB} G.E.~Brown and M. Bolsterli, Phys. Rev. Lett. {\bf 3}, 472(1959)
\bibitem{SW} B.~Schwesinger and J.~Wambach, Nucl. Phys. {\bf A426},
253(1984)
\bibitem{Izra} F.M.~Izrailev, Phys. Rep. {\bf 196}, 299(1990)
\bibitem{FLW} M.~Feingold, D.M.~Leitner and M.~Wilkinson,
Phys. Rev. Lett. {\bf 66}, 986(1991)
\bibitem{FGGK} V.V.~Flambaum, A.A.~Gribakina, G.F. Gribakin and M.G.~Kozlov,
Phys. Rev. {\bf A50}, 267(1994)
\bibitem{Wies} K.~Wiesenfeld and F.~Moss, Nature {\bf 373}, 33(1995); \\
A.R.~Bulsara and L.~Gammaitoni, Physics Today, 39(March 1996) 
\end{references}
\end{document}